\documentclass[aps,prl,twocolumn]{revtex4}
\usepackage{epsfig}
\def\V2O3{$\mathrm{V_2O_3}$}
\begin{document}

\begin{abstract}
Thin films of \V2O3 have been grown on Au(111) and W(110). It is possible to prepare two different surface terminations: the first one is vanadium terminated whereas the second one exhibits additional oxygen atoms, forming vanadyl groups with the surface vanadium atoms. The electronic structure was studied for both terminations by photoelectron spectroscopy. While the first surface is metallic at room temperature like \V2O3 bulk, the second surface with the vanadyl groups shows a gap at the Fermi level.
\end{abstract}

\title{V$_2$O$_3$(0001) on Au(111) and W(110): Metal to Insulator Transition Induced by Surface Termination}
\author{A.-C. Dupuis
\footnote{corresponding author,
        e-mail: ac.dupuis@web.de} 
}
\affiliation{Fritz-Haber-Institut der Max-Planck-Gesellschaft,
Faradyaweg 4-6, 14195 Berlin, Germany}
\maketitle

\section{introduction}
Band theory's prediction whether a material at 0K is metallic or insulator is based on the filling of electronic bands: For insulators, the highest filled band is completely filled ; for metals, it is partially filled. This band picture is based on noninteracting or weakly interacting electron systems. However, many transition metal oxides with a partially d electrons band show insulator behavior. The breakdown of the band theory for these compounds evidences the importance of electron-electron correlations. V$_2$O$_3$, with its very rich phase diagram, is a good example to
demonstrate the complexity of transition metal oxides. Of
particular interest is its paramagnetic metallic (PM) to
antiferromagnetic insulator (AFI) phase transition at 150 K. This
metal insulator transition (MIT) is generally believed to result from electron-electron correlations(Mott-Hubbard transition). However, despite the large number of experimental and theoretical works done on \V2O3, the understanding of the transition and the nature of the ground state are still controversial. 

Most of the experimental studies on V$_2$O$_3$ reported in the
literature were done on single crystals or powder materials and only
a few on thin or ultra thin films. We are able to grow a thin
V$_2$O$_3$ film which exhibits the same geometric and electronic
structure as a single crystal. We showed that two terminations for the \V2O3(0001) surface exist \cite{LACgrowth}. The difference between both surface terminations is the presence or absence of additional oxygen atoms at the surface. These oxygen atoms form a double bond with the surface vanadium atoms, thus creating vanadyl groups. We observed a
dramatic dependence of the electronic structure on the surface
termination. Indeed, our photoemission results evidence a MIT induced by the formation of the vanadyl
species at the surface.

\section{Experimental}
Three different UHV systems were employed for this work. Angle
resolved UV photoelectron spectra (ARUPS) were taken in the first
one using a VSW ARIES spectrometer with a rotatable electron
analyzer and a Specs helium discharge lamp as source for UV
radiation. 
X-ray photoemission spectroscopy
(XPS) and near edge X-ray absorption spectroscopy (NEXAFS)
measurements were performed in the second system at the BESSY II
synchrotron radiation source in Berlin. Infrared absorption
spectra (IRAS) were taken in the third system with a modified
Mattson RS-1 FTIR spectrometer. The measurement temperature was 300 K for all the spectra shown
here.

Au(111) substrate was cleaned in UHV by alternating cycles of
Argon sputtering and annealing at 1150 K. To remove carbon from
W(110) we repeated the usual cleaning cycle consisting of heating
the sample first to 1800 K in 10$^{-6}$ mbar of oxygen during a
few minutes and then to 2300 K without oxygen. The
V$_2$O$_3$(0001) film was prepared for both substrates by
evaporation of metallic vanadium in an oxygen atmosphere
(10$^{-7}$ mbar), followed by annealing at 700 K in 5.10$^{-8}$
mbar of oxygen. The vanadium oxide so obtained is stable under UHV
conditions (10$^{-10}$ mbar) up to high temperatures (at least
1050 K). However, heating to higher temperatures than 800 K leads
to diffusion of gold substrate atoms towards the surface. This diffusion
phenomenon does not seem to occur with the tungsten substrate.

\section{Results}
V$_2$O$_3$ has corundum structure like Al$_2$O$_3$, i.e. the
oxygen atoms form a hexagonal lattice and two third of its
octahedral sites are occupied by vanadium atoms. The good quality
of the LEED pattern obtained for a 30 \AA \ thick film on both
substrates provides evidence for satisfactory epitaxial growth. We
controlled the stoichiometry of the V$_2$O$_3$ film with XPS and
NEXAFS. We can conclude that the film we obtained is identical in
terms of its stoichiometry and electronic structure to a single
crystal of V$_2$O$_3$ (see \cite{LACgrowth} for details about
the growth of the thin film).

\begin{figure}
\centerline{
             \epsfig{file=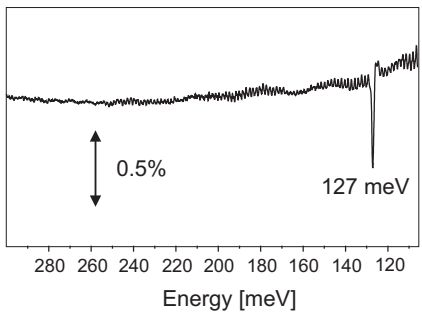,width=5.5cm}
             }  
  \caption{IRAS: spectrum for -V=O termination divided by spectrum for -V
termination. The absorption peak corresponds to the V-O stretching
vibration of the vanadyl groups}
   \label{f:IRAS_VO}
\end{figure}

The first surface termination was obtained immediately after the
normal preparation process described above. The second one was
obtained from the first one by heating the sample up to 600 K with
electron bombardment for a short time (few seconds). Note that
this process is reversible : annealing the second surface leads to
the first one again. We performed IRAS measurements for both
terminations. A commonly used method by IRAS in surface science is to take a first spectrum for a reference surface and a second one for the surface to analyze and then make the ratio between both sets of data. By adsorption experiments, for example, the reference surface is the clean surface and the surface to analyze the covered surface. It allows one to observe the absorption features of only the adsorbate. We used this method here, setting the data of the second terminated surface as reference to analyze the data of the first surface. The
result, shown in Figure \ref{f:IRAS_VO}, exhibits only one absorption feature at
127 meV (=1023 cm$^{-1}$).

A feature at the same energy was previously observed for
$\mathrm{V_2O_5}$(001) by high resolution electron energy lost spectroscopy (HREELS) and assigned to the V-O
stretching vibration of vanadyl groups (V=O) \cite{L136}. Netzer
et al. showed also the existence of vanadyl groups in
V$_2$O$_3$(0001)/Pd(111) \cite{L138}. We therefore conclude that the difference between both surface terminations is only the presence of vanadyl species on the first surface. In the following we will
call the first surface termination with the vanadyl species the
-V=O termination and the other one, where the V=O groups have been
removed, the -V termination.

We used UPS to study the surface electronic structure of the V$_2$O$_3$ film
with both -V and -V=O terminations.
\begin{figure}[htbp]
 \centerline{
             \epsfig{file=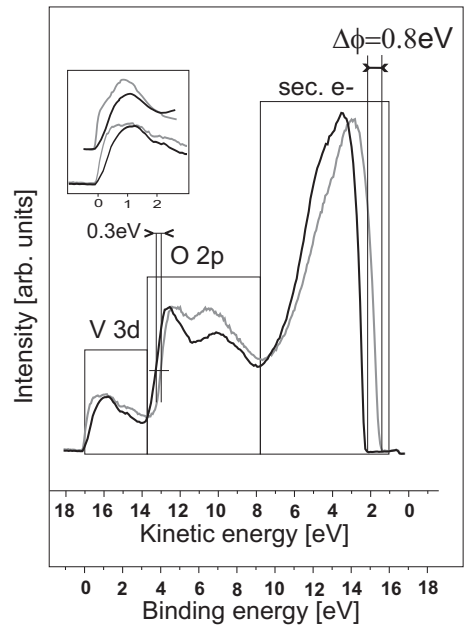, width=6cm}
             }  
  \caption{UPS (HeI) on V$_2$O$_3$(0001)/Au(111) for the -V (grey) and -V=O
  (black) terminations at normal emission. The difference of the low energy
  cutoff corresponds to the difference of the work function between the two terminations.
  Inset : comparison of our spectra (below) with spectra from \cite{LGoering} of V$_2$O$_3$ single crystal
  above and below the transition temperature (above) in the V 3d emission region}
  \label{f:UPS_HeI}
  \end{figure}
We observe mainly four differences between the spectra of the -V=O termination and the spectra of the -V termination, as will be discussed in detail below. First, there is an increase of the work function. Second, the O 2p band exhibits a shift towards lower binding energies. Third, the spectra exhibit additional features within the O 2p band. And at last, the spectra show a decrease of emission at the Fermi energy.
\begin{figure*}[htbp]
  \centerline{
             \epsfig{file=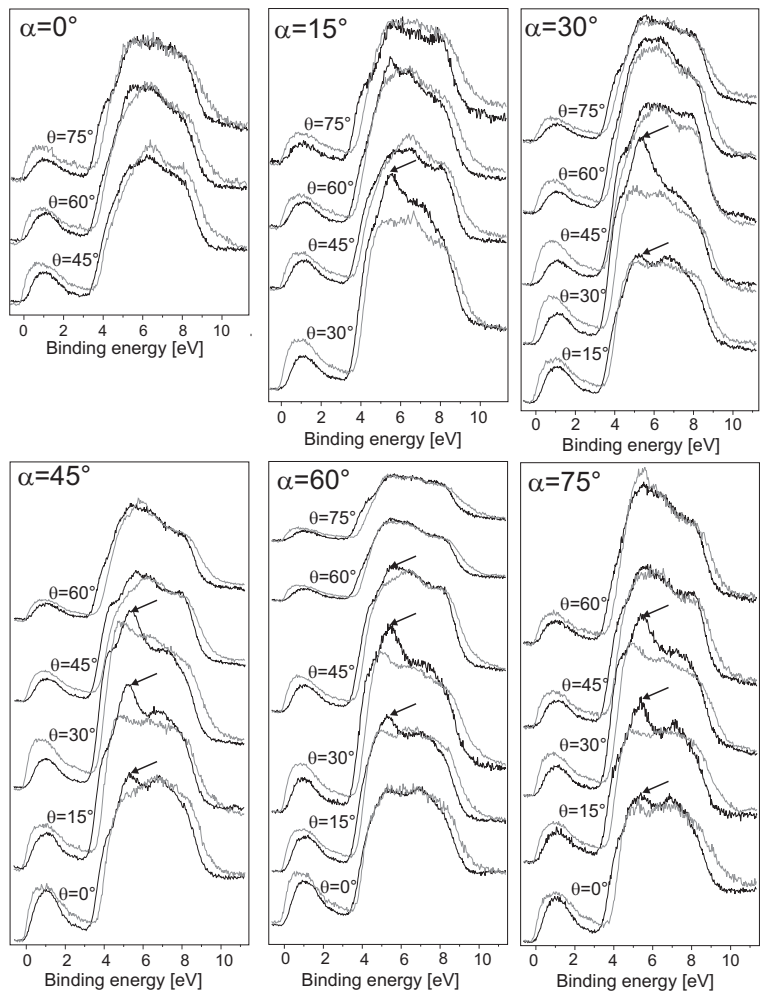, width=10cm}
             }  
  \caption{ARUPS on V$_2$O$_3$(0001)/Au(111) for the -V (grey) and -V=O
  (black) terminations - $\alpha$ is the light angle of incidence and $\theta$ the angle of detection}
  \label{f:UPS_term}
  \end{figure*}
We show in Figure \ref{f:UPS_HeI} a He I spectrum for each
termination at room temperature. The light angle of incidence was 45$\mathrm{^o}$ with respect to the surface normal; the electrons were detected in the direction normal to the surface.
One can distinguish three bands
of emission. The first and least intense one between the binding energies ($E_B$) 0
(corresponding to the Fermi energy E$_\mathrm{F}$) and 3.5eV
originates from the V 3d bands. The second in the middle from
about 3.5 to 9.5eV corresponds to the O 2p band. The last one
arises from secondary electrons. The assignment for the first two
features corresponds to an interpretation in terms of a simple
ionic picture but band structure calculations
\cite{L21,L56,LHermann} and theoretical analysis of core level
spectra \cite{L27} show evidence for a strong V 3d - O 2p
hybridization. The low energy cutoff gives us information about
the work function of the material. We find a difference of 0.8 eV
between the two terminations. This is not surprising since the
vanadyl species form a dipole with its negative end outwards,
which increases the work function. The O 2p band of the -V=O
terminated surface exhibits a shift of about 0.3 eV towards lower
binding energies relative to the -V terminated surface. This shift
is indicated on Figure \ref{f:UPS_HeI}.

We investigated more thoroughly the differences in the electronic
structure between both surfaces using HeII angle resolved
photoelectron spectroscopy. We performed a measurement series for
6 incidence angles and 6 detection angles (0, 15$^\mathrm{o}$, 30$^\mathrm{o}$,
45$^\mathrm{o}$, 60$^\mathrm{o}$ and 75$^\mathrm{o}$), the detector being positioned in the incidence
plane. The results are shown in Figure \ref{f:UPS_term}.

For geometrical reasons only 28 different combinations of incident
light and emission angles were available for measurement. The
spectra were normalized to the background signal above
E$_\mathrm{F}$. No background has been removed. 

As for the spectra
of Figure \ref{f:UPS_HeI}, a shift of the O 2p band between both
terminations is observed for each measurement geometry. A shift of the O 2p band
away from E$_\mathrm{F}$ has been reported for Cr-doped V$_2$O$_3$
in \cite{L70} when the system becomes insulating. This shift is in
the opposite direction than the one we observed. This suggests that there is no MIT in the bulk but that the effect observed concerns only the surface.

The spectra of the -V=O surface exhibit an additional spectral feature
at about 5.5 eV, showing strong angular dependence. This feature
is indicated on Figure \ref{f:UPS_term} by arrows. It is assigned to
O 2p electrons of the vanadyl species.  

For the -V=O termination, all spectra of Figure \ref{f:UPS_term}
exhibit a decrease of the emission intensity from the V 3d band
and a narrowing of this band. More precisely, there is no spectral
weight at the Fermi level, demonstrating the opening of a gap in
the surface region. The opening of a similar gap was observed at the PM-AFI transition of pure V$_2$O$_3$ 
pure V$_2$O$_3$ \cite{L25,L66,L102,LGoering,L81} and of Cr-doped
V$_2$O$_3$ single crystals \cite{L70}. We show for comparison in the inset of Figure
\ref{f:UPS_HeI} He I spectra in the V 3d region from
\cite{LGoering} obtained on V$_2$O$_3$ single crystal above and
below the transition temperature. Their spectra show a quite
similar shape as ours. This gives evidence that the formation of vanadyl
groups on the vanadium terminated surface therefore induces a MIT at the surface.

\section{Discussion}
Formation of vanadyl groups on the V$_2$O$_3$ surface implies a
change in the occupied states in the V 3d band. V 3d
orbitals are supposed to mix with O 2p orbitals to form the double
V-O bond of the vanadyl species. The V 3d band will then be partly
depleted because V 3d electrons will be involved in the vanadyl
bonding orbitals. In a metal, a simple decrease of the electronic density in the V 3d band should only lead to a  "repositioning" of the Fermi level
relative to the V 3d band. One would find a similar emission intensity at the Fermi level for both terminations. The difference for both terminations would simply be a narrowing of the V 3d band for the -V=O termination and a shift of the O 2p band relative to the Fermi level. Therefore, the decrease of emission at E$_\mathrm{F}$ for the -V=O surface
evidences an opening of a gap. The band structure calculation performed by Mattheiss et al. \cite{L56} indicates that the partially filled bands at the Fermi surface of the metallic phase of bulk \V2O3 involve all 5 $\mathrm{a_{1g}}$, $\mathrm{e_g^\pi}$ and $\mathrm{e_g^\sigma}$ sates. Thus, the gap we observed for the -V=O surface cannot result from a band effect but has to be interpreted as a correlation gap as it is the case for the low temperature phase of bulk \V2O3.

The Mott transition can be characterized in photoemission by a decrease of emission at the Fermi level. In the framework of the so-called dynamical mean-field theory (DMFT), this change in emission from the d band is believed to correspond to a transfer of spectral weight from the region near the Fermi level to the upper and lower Hubbard bands. The DMFT has been used to explain photoemission spectra of $\mathrm{3d^1}$ compounds \cite{L120,L221,L225,L229}. These compounds show in the 3d emission region two features: the first one, just below $\mathrm{E_F}$, s assigned to the coherent part of the single-particle spectral function, and the second one to its incoherent part (corresponding to the lower Hubbard band). The coherent part seems to correspond in this model to the density of states and is attributed to itinerant d-band states or quasi-particle excitations \cite{L221}. The changes in the emission from the V 3d band between both terminations reveal actually two effects: an opening of a correlation gap and a decrease of the electronic density in the V 3d band for the -V=O termination. Since the latter implies a decrease of the emission from the V 3d band, one may have some difficulty to interpret the former along the line of the DMFT, i.e. in terms of transfer of spectral weight.

The MIT we observed when vanadyl groups form on the surface is a very surprising effect. In a first approximation, formation of vanadyl groups can be interpreted as a doping of holes. It has been observed on ternary compounds \cite{L221} and also theoretically shown \cite{L149a} that doping a Mott insulator with carriers (holes or electrons) leads to a transfer of spectral weight to the Fermi level. Formation of vanadyl groups should therefore favor the metallic phase and not induce a MIT, as we experimentally observed. Theoretical treatments had therefore to be performed to understand the observed phenomenon properly.

In conclusion, we studied the electronic structure of thin films of \V2O3(0001) with UV photoelectron spectroscopy (UPS). The results show a decrease of spectral weight at the Fermi level when vanadyl groups form on the vanadium terminated surface, evidencing a Mott transition.

\section{Acknowledgment}
We thank the Fritz-Haber-Institut for having supported this work and Silke Biermann for interesting discussions.

\bibliographystyle{unsrt}

\end{document}